\documentstyle[epsf]{elsart}

\begin{document}
\begin{frontmatter}

\title{Localization in non-Hermitian quantum mechanics\\
and flux-line pinning in superconductors\thanksref{talk}}
\thanks[talk]{An invited talk at StatPhys-Taipei-1997 
(Taiwan, August 1997).}

\author{Naomichi Hatano\thanksref{e-mail}}
\thanks[e-mail]{e-mail: hatano@viking.lanl.gov}
\address{Theoretical Division, Los Alamos National Laboratory,\\
Los Alamos, New Mexico 87545, USA}

\begin{abstract}
A recent development in studies of 
random non-Hermitian quantum systems is reviewed.
Delocalization was found to occur under a sufficiently large constant 
imaginary vector potential even in one and two dimensions.
The phenomenon has a physical realization as flux-line depinning
in type-II superconductors.
Relations between the delocalization transition and the complex energy
spectrum of the non-Hermitian systems are described.
Analytical and numerical results obtained for a non-Hermitian Anderson 
model are shown.
\end{abstract}


\end{frontmatter}

\section{Introduction}

The main purpose of the present paper is to review a new development
in studies of non-Hermite operators.
The review is mostly based on the work by the present author in 
collaboration with Nelson~\cite{Hatano96,Hatano97} and the works
following 
it~\cite{Shnerb97,Efetov97,Feinberg97a,Janik97c,Brouwer97,%
Feinberg97c,Goldsheid97,Brezin97,Nelson97,Feinberg97d,Janik97e,Zee97}.
In addition, I report
a new numerical result for a non-Hermitian ladder system.

Non-Hermite operators appear frequently in 
dynamical systems as Liouville operators.
Although less frequently, they also appear in the context of quantum 
mechanics as Hamiltonian operators in the Schr\"{o}dinger equation.
A well-known example is the optical potential, which is a
complex scalar potential that effectively describes multiple scattering
and absorption.
A non-Hermite Hamiltonian 
can also emerge when one relates a $(d+1)$-dimensional classical 
statistical system to a $d$-dimensional quantum system by path-integral 
scheme or the Suzuki-Trotter transformation~\cite{Suzuki76}.
As a classic example, McCoy and Wu~\cite{McCoy68} 
showed that equilibrium classical statistical mechanics of a
two-dimensional asymmetric vertex model can be described
by imaginary-time quantum dynamics of a non-Hermitian $XXZ$ spin chain.
In this case, the non-Hermiticity of the spin chain is originated in
an external field that generates a diagonal flow of edge spins of 
the vertex model.

The new development reviewed here resulted from introduction of 
{\em quenched randomness} into non-Hermitian quantum systems.
A delocalization phenomenon was found for an especially simple class of 
random non-Hermite Hamiltonians even in one and two dimensions.
The Hamiltonian contains a constant {\em imaginary} vector potential 
and a real {\em random} scalar potential.
As the imaginary vector potential increases, 
all of originally localized eigenfunctions get delocalized one by one.
One of the remarkable features is that
a complex eigenvalue of the Hamiltonian indicates the delocalization
of the corresponding eigenfunction.
Thus we can study the delocalization phenomenon simply by investigating
the energy spectrum of the non-Hermite Hamiltonian.

The delocalization phenomenon has a physical realization as 
flux-line depinning in type-II superconductors.
Basic correspondence between the delocalization and the depinning
is described in the next section.
In Section~\ref{spectra}, I discuss the complex-energy spectrum 
of the random non-Hermitian system, using numerical data
for a non-Hermitian Anderson model.
A new result for a non-Hermitian ladder system is also reported.
Section~\ref{Meissner} presents an interesting application of Mott's 
variable-range hopping to the non-Hermitian quantum mechanics.

Non-Hermite matrices with randomness have recently attracted much attention
from other viewpoints as well.
The spectrum of the Fokker-Planck operator with a random velocity field
has been studied in Refs.~\cite{Miller96,Chalker97}.
Non-Hermitian random matrix theory has seen great progress
recently~\cite{Feinberg97a,Janik97e,%
Fyodorov96,Fyodorov97a,Fyodorov97b,Fyodorov97c,%
Janik97a,Janik97b,Janik97d,Feinberg97b}.
The present review, however, does not cover these topics.

\section{Non-Hermitian quantum mechanics and flux lines
in superconductors}

\subsection{Random non-Hermite Hamiltonian and an elastic string
in a random washboard potential}

A typical Hamiltonian which I treat here is
\begin{equation}\label{12020}
{\cal H}
\equiv\frac{(\vec{p}+\mathrm{i}\vec{g})^{2}}{2m}
+V(\vec{x}),
\end{equation}
where $\vec{p}=(\hbar/\mathrm{i})\partial/\partial \vec{x}$ 
is the momentum operator, $\vec{g}$ is a constant real vector
referred to as a non-Hermitian field, and $V$ is a random potential.
A lattice version of the above Hamiltonian is given by
a non-Hermitian Anderson model on a hypercubic lattice,
\begin{equation}\label{41010}
{\cal H}\equiv 
\sum_{\vec{x}} \left[
-\frac{t}{2}
\sum_{\nu=1}^{d} \left(
\e^{\vec{g}\cdot\vec{e}_{\nu}/\hbar}
\Bigl|       \vec{x}+\vec{e}_{\nu} \Bigr\rangle
\Bigl\langle \vec{x}               \Bigr|
+\e^{-\vec{g}\cdot\vec{e}_{\nu}/\hbar}
\Bigl|       \vec{x}               \Bigr\rangle
\Bigl\langle \vec{x}+\vec{e}_{\nu} \Bigr|
\right)
+
V_{\vec{x}} 
\Bigl|       \vec{x} \Bigr\rangle
\Bigl\langle \vec{x} \Bigr|
\right],
\end{equation}
where $t$ is the hopping amplitude, 
the vectors $\{\vec{e}_{\nu}\}$
are the unit lattice vectors, and
$V_{\vec{x}}$ is an on-site random potential
following a probability distribution $P(V_{\vec{x}})$.
Periodic boundary conditions are applied to wave functions
in both cases.
Note that the non-Hermitian field $\vec{g}$ plays a role of an 
{\it imaginary} vector potential.

In the Hermitian case $\vec{g}=\vec{0}$, Eqs.~(\ref{12020})
and~(\ref{41010}) are reduced
to the standard Hamiltonians for the Anderson localization.
It is widely accepted for $\vec{g}=\vec{0}$ that 
all eigenfunctions are localized in one and two dimensions.
The present author and Nelson recently found~\cite{Hatano96,Hatano97} 
that all of the localized eigenfunctions get delocalized one by one
as the non-Hermitian field $\vec{g}$ is increased
and that appearance of complex eigenvalues indicates
the delocalization transition.

As is exemplified in the study of McCoy and Wu~\cite{McCoy68},
a non-Hermite Hamiltonian can have physical relevance when it
is mapped to a classical statistical-mechanical system with
path-integral mapping.
By identifying the imaginary-time-evolution operator
$\e^{-\Delta\tau{\cal H}}$ of the Hamiltonian~(\ref{12020})
as the transfer matrix of a classical system, 
we can transform~\cite{Feynman65,Schulman81} the matrix element
of the time-evolution operator between the initial and final vectors,
\begin{equation}\label{11020-05}
{\cal Z}\equiv
\left\langle\psi^{\mathrm{f}}\Bigm|
\e^{-L_{\tau}{\cal H}/\hbar}\Bigm|
\psi^{\mathrm{i}}\right\rangle,
\end{equation}
into the partition function of an elastic string
in a $(d+1)$-dimensional space,
\begin{equation}\label{11020-1}
{\cal Z}=\int{\cal D}\vec{x}
\e^{-{E_{\mathrm{cl}}[\vec{x}(\tau)]}/\hbar}
\end{equation}
with the energy of the string given by
\begin{equation}\label{00000}
E_{\mathrm{cl}}[\vec{x}(\tau)]
\equiv\int_{0}^{L_{\tau}}\d\tau
\left[
\frac{m}{2}
\left(\frac{\d\vec{x}}{\d\tau}\right)^{2}
-\vec{g}\cdot\frac{\d\vec{x}}{\d\tau}
+V(\vec{x})
\right].
\end{equation}
The first term of the energy describes the elasticity of the string.
(See Appendix for details of the above transformation.)

The energy~(\ref{00000}) of the elastic string is re-interpretation of 
the imaginary-time action of the quantum particle.
The imaginary-time axis $\tau$ of the quantum system is identified
as an additional spatial axis of the classical system. Hence
the imaginary time $L_{\tau}$ becomes the system size of the classical 
system in the $\tau$ direction.
The world line of the quantum particle, $\vec{x}(\tau)$, is 
identified as a spatial configuration of the classical elastic string 
subject to thermal fluctuations.
The temperature of the classical system is given by the Planck parameter
$\hbar$.
The integral $\int{\cal D}\vec{x}$ denotes the summation over all
possible configurations of the world line, or the elastic string.

\begin{figure}
\epsfxsize=8cm
\epsfbox{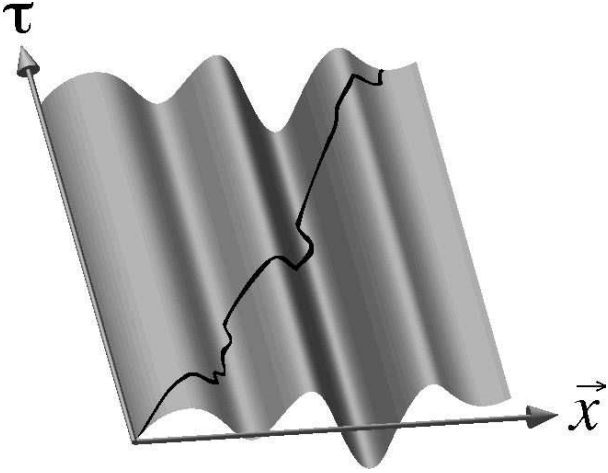}
\caption{
An elastic string in a random washboard potential is subject 
to thermal fluctuations. 
The energy of the string is given by Eq.~(\protect\ref{00000}).
Through path-integral mapping, statistical mechanics of the system
is equivalent to quantum dynamics of the Hamiltonian~(\protect\ref{12020}).
}
\label{fig1}
\end{figure}

Note that the random potential does not depend on $\tau$.
Hence the elastic string is put on a \lq\lq random washboard'' 
potential (Fig.~\ref{fig1}). 
The potential tries to trap the elastic string in 
particularly deep valleys and thereby
to align the string along the $\tau$ direction.
On the other hand, the second term of the energy~(\ref{00000}),
which comes from the non-Hermitian field $\vec{g}$,
tries to tilt the string away from the $\tau$ direction as is explained below.
(In fact, if the random potential is not present, the energy is optimized when 
the string is tilted by the angle $\tan^{-1}(g/m)$.)
The competition between the above two effects results in a pinning-depinning 
transition of the string from the random washboard potential.
This in turn indicates a localization-delocalization transition of
the quantum particle subject to the non-Hermitian field and the random
potential.

The reason why the non-Hermitian field $\vec{g}$ in the quantum system
has the effect of tilting the elastic string may be explained in the
following way.
A vector potential $\vec{A}$ generally induces 
a current in a quantum system because of the coupling $\vec{p}\cdot\vec{A}$
in the Hamiltonian.
The current would be expressed by the world lines of quantum particles 
running diagonally in the {\it real-time}-space.
Since the present non-Hermitian field plays a role of an {\it imaginary}
vector potential, it induces an {\em imaginary current}.
The imaginary current is expressed by tilted world lines in the
{\em imaginary-time}-space.
Hence the field $\vec{g}$ tries to tilt the world line, or the elastic string.

\subsection{Flux-line pinning in type-II superconductors}

The situation described by Eq.~(\ref{00000}) is realized in type-II 
superconductors with extended defects.
In a harmonic approximation, Nelson and Vinokur~\cite{Nelson93} derived
a phenomenological Hamiltonian of a flux line
in a superconductor with extended defects.
The Hamiltonian takes the form of Eq.~(\ref{00000}) with $V(\vec{x})$ 
denoting the pinning potential due to the defects.

\begin{figure}
\epsfxsize=8cm
\epsfbox{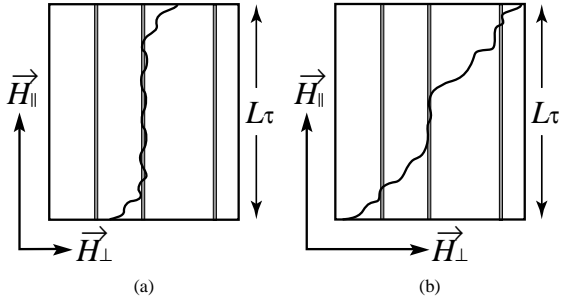}
\caption{
Flux-line depinning due to tilt of the external magnetic field.
(a) When the transverse component of the magnetic field is small,
the flux line is pinned by a columnar defect in the bulk of the 
superconductor, although it is deflected near the surfaces.
(b) For a larger $\vec{H}_{\perp}$, the flux line is 
depinned from the defect.
}
\label{fig2}
\end{figure}

When an electric
current is applied to a pure sample of a type-II superconductor in the mixed
phase, magnetic flux lines penetrating the sample are moved by 
electromagnetic forces, dissipate energy, and thus destroy the 
superconductivity.
Randomly located (but mutually parallel)
extended defects such as columnar defects (typically created by 
bombardment of heavy ions) and twin boundaries (planer defects in the
anisotropic YBCO) pin the flux lines efficiently as long as the flux lines
are almost parallel to the defects~\cite{Nelson93,Civale91,Budhani92}.
When the external magnetic field is tilted away from the defects,
but its transverse component $\vec{H}_{\perp}$ is still small,
we may expect that the bulk part of the flux line remains pinned 
(Fig.~\ref{fig2}(a)).
This is referred to as the {\em transverse Meissner effect}~\cite{Nelson93},
because the system in this region exhibits perfect bulk diamagnetism 
in the transverse direction.
When one increases the tilt angle (Fig.~\ref{fig2}(b)), a depinning 
transition occurs at a certain strength of the transverse magnetic field, 
namely $H_{\perp\mathrm{c}}$.
(See Section~\ref{Meissner} for further details.)

The above depinning transition can be explained in a clear-cut way
in terms of delocalization in the random non-Hermitian 
quantum mechanics~\cite{Hatano96,Hatano97}.
The depinning of the flux line corresponds to delocalization of 
the relevant wave function of the quantum system;
see Table~\ref{tab1} for other correspondence.
\begin{table}
\caption{Correspondence between the random non-Hermitian system and 
the flux-line system.}
\label{tab1}
\begin{center}
\begin{tabular}{ll}
\hline
Random non-Hermitian system & Flux-line system \\ 
\hline\hline
$d$-dimensional space & $(d+1)$-dimensional space \\ 
\qquad and the imaginary time & \\
Quantum fluctuation & Thermal fluctuation \\
World line & Flux line \\
Non-Hermitian field $\vec{g}$ 
& Transverse field $\vec{H}_{\perp}$ \\
Random potential $V(\vec{x})$ 
& Randomly located extended defects \\
Localization & Pinning \\
Delocalization & Depinning \\
Real eigenvalue & Transverse Meissner effect \\
Complex eigenvalue in a periodic system & Helical structure of a flux line \\
Imaginary current of a quantum particle & Tilt of a flux line \\
\hline
\end{tabular}
\end{center}
\end{table}

Although it would be possible to describe the depinning within the
framework of classical statistical mechanics, one can take advantage, in 
the non-Hermitian approach, of abundant results available concerning
localization in Hermitian random systems.
It is also often practically easier to solve a 
Schr\"{o}dinger equation 
than to treat the same problem in the path-integral framework.

There are many other physical realizations 
of the above random non-Hermitian system.
Efetov~\cite{Efetov97} discussed the problem from the viewpoint of
{\em directed} quantum chaos.
Nelson and Shnerb found an interesting application to population 
biology~\cite{Nelson97}.
In an independent work, Chen {\it et al}.~\cite{Chen96} 
employed the same technique to study
sliding of charge-density waves in disordered systems.
In the following, I concentrate on the flux-line analogy.

\section{Complex eigenvalues and delocalization}
\label{spectra}

\subsection{Energy spectrum in one dimension}

\begin{figure}
\epsfxsize=8cm
\epsfbox{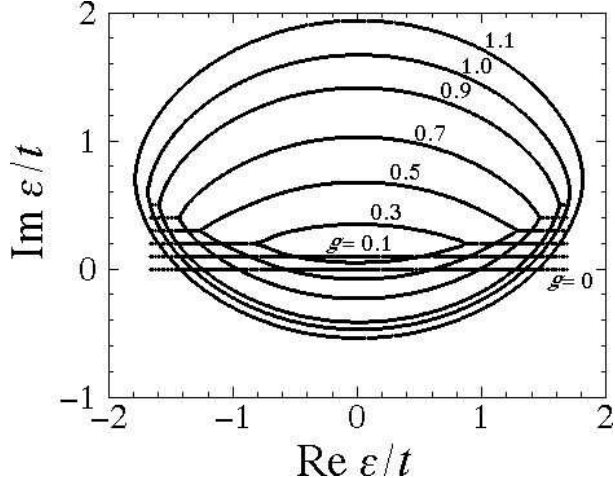}
\caption{
The energy spectrum of the one-dimensional non-Hermitian 
model~(\protect\ref{41010}) with 1000 sites.
Each eigenstate $\varepsilon$ is marked by a tiny cross 
in the complex energy plane.
Plots for different values of $g$ are offset for clarity.
The random potential at each site was chosen from a 
box distribution over the range $[-t,t]$.
The same realization of the random potential $\{V_{x}\}$ 
was used for all plots here.
A complex eigenvalue indicates that the eigenstate is 
delocalized and the corresponding flux line is depinned.
}
\label{fig3}
\end{figure}

I now present an example of the energy spectrum
of the random non-Hermitian system~\cite{Hatano96,Hatano97}.
Figure~\ref{fig3} shows numerical results for 
the lattice Hamiltonian~(\ref{41010}) in one dimension
with 1000 sites.
All energy eigenvalues are of course real for the Hermitian case $g=0$.
For weak $g$, {\it e.g.}\ $g/\hbar=0.1$ in the case of Fig.~\ref{fig3}, 
all the eigenvalues are still real
despite the fact that the Hamiltonian is non-Hermite.
As we increase $g$, complex eigenvalues appear in the middle of
the energy band and form a bubble.
(The spectrum has an inversion symmetry with respect to the 
real energy axis, because the Hamiltonian~(\ref{41010}) is a real matrix.)
The region of complex eigenvalues expands towards the band edges
as $g$ is increased,
and the whole spectrum eventually becomes almost elliptic.
In fact, the spectrum exactly falls onto an ellipse 
if $V_{x}\equiv0$~\cite{Hatano96,Hatano97}:
\begin{equation}\label{2004}
\left(\frac{\mathop{\mathrm{Re}}\varepsilon}{\cosh(g/\hbar)}\right)^{2}
+\left(\frac{\mathop{\mathrm{Im}}\varepsilon}{\sinh(g/\hbar)}\right)^{2}
=t^{2}.
\end{equation}

Analytic forms of the random-averaged energy spectrum in one dimension 
have been obtained for weak disorder and weak $g$~\cite{Brouwer97}
and for general values of $g$ with the Lorentzian random 
distribution~\cite{Goldsheid97,Brezin97}.
By neglecting higher moments of the random distribution of 
the on-site potential $\{V_{x}\}$,
Brouwer {\it et al.}\ obtained an approximate shape of the bubble after
averaging, in the form
\begin{equation}\label{2000}
|\mathop{\mathrm{Im}}\varepsilon|=
\frac{|g|}{\hbar}\sqrt{t^{2}-(\mathop{\mathrm{Re}}\varepsilon)^{2}}
-\frac{\Delta^{2}}{2\sqrt{t^{2}-(\mathop{\mathrm{Re}}\varepsilon)^{2}}}
\end{equation}
for $|\mathop{\mathrm{Re}}\varepsilon|<\varepsilon_{\mathrm{c}}$ and for
small $|g|$,
where $\Delta^{2}\equiv\langle V_{x}^{2} \rangle$ 
is the second moment of the random distribution and
\begin{equation}\label{2001}
\varepsilon_{\mathrm{c}}\equiv\sqrt{t^{2}-\frac{\hbar\Delta^{2}}{2|g|}}.
\end{equation}
The value of $\varepsilon_{\mathrm{c}}$
indicates the region of the bubble of complex eigenvalues and in fact 
is a \lq\lq mobility edge'' as discussed below.
Note that the bubble does not exist for $|g|<\hbar\Delta^{2}/(2t^{2})$.

For the Lorentzian distribution of the potential,
$P(V_{x})=\pi^{-1}\gamma/(V_{x}^{2}+\gamma^{2})$, 
whose second moment is diverging,
the exact shape of the bubble after averaging was
obtained for general values of $g$~\cite{Goldsheid97,Brezin97}:
\begin{equation}\label{2005}
\left(\frac{\mathop{\mathrm{Re}}\varepsilon}{\cosh(g/\hbar)}\right)^{2}
+\left(\frac{|\mathop{\mathrm{Im}}\varepsilon|+\gamma}%
{\sinh(g/\hbar)}\right)^{2}=t^{2}
\end{equation}
for $|\mathop{\mathrm{Re}}\varepsilon|<\varepsilon_{\mathrm{c}}$
with the \lq\lq mobility edge''
\begin{equation}
\label{2006}
\varepsilon_{\mathrm{c}}\equiv\cosh(g/\hbar)
\sqrt{t^{2}-\gamma^{2}{\mathop{\mathrm{sech}}}^{2}(g/\hbar)}.
\end{equation}
The form~(\ref{2005}) is remarkably simple;
the upper and lower halves of the elliptic spectrum of the pure case,
Eq.~(\ref{2004}), are
squeezed towards the real energy axis translationally 
by the distance $\gamma$, preserving the shape of the arcs.
Eigenvalues that lost the support of the arcs become real and are
distributed over the whole real axis except for the region of the bubble.
The bubble entirely vanishes for $|g|<\hbar\sinh^{-1}(\gamma/t)$.

Another approximate expression of the mobility edge $\varepsilon_{\mathrm{c}}$
has been obtained for general random distribution~\cite{Janik97c,Janik97e}.
For the Lorentzian randomness, 
this expression is reduced to the exact result~(\ref{2006}).

\subsection{Delocalization and complex eigenvalues}

\begin{figure}
\epsfysize=7cm
\epsfbox{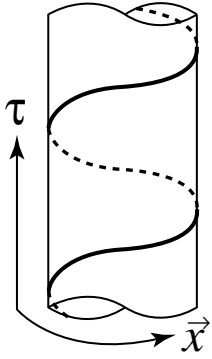}
\caption{
The world line (the thick line)
of a current-carrying particle forms a helix
when periodic boundary conditions are imposed in the space direction.
Hence periodicity appears in the imaginary-time direction.
The periodicity is described by an oscillatory factor 
$\e^{\mathrm{i}\tau\mathop{\mathrm{Im}}\varepsilon}$.
}
\label{fig4}
\end{figure}

In the following, I explain the correspondence that is listed in
Table~\ref{tab2}.
\begin{table}
\caption{
The delocalization criterion of the random non-Hermitian system.
A wave function of the form $\psi\sim\e^{-\kappa|\vec{x}|}$ for 
$\vec{g}=\vec{0}$
and $|\vec{x}|\to\infty$ gets delocalized when $|\vec{g}|$ exceeds
$\hbar\kappa$, and acquires a complex eigenvalue at the same time.
}
\label{tab2}
\begin{center}
\begin{tabular}{ccc}\hline
non-Hermitian field $\vec{g}$ & wave function & eigenvalue \\ \hline\hline
$|\vec{g}|<\hbar\kappa$            & localized     & real       \\
$|\vec{g}|>\hbar\kappa$            & delocalized   & complex    \\ \hline
\end{tabular}
\end{center}
\end{table}
I first argue that a complex eigenvalue indicates a delocalized wave function.
Consider the imaginary-time dynamics of a wave function
\begin{equation}
\label{1}
\psi(\vec{x};\tau)=\psi(\vec{x})\e^{-\tau \varepsilon}\propto
\e^{-\mathrm{i}\tau\mathop{\mathrm{Im}}\varepsilon},
\end{equation}
where $\varepsilon$ is the eigenvalue of the eigenfunction $\psi(\vec{x})$.
The above equation shows that the imaginary part of the eigenvalue, if
it exists, gives rise to an oscillatory behavior of the system in the
imaginary-time direction.
In fact, the oscillatory behavior comes from the world line
wrapping around the system as a helix (Fig.~\ref{fig4}).
Note here that 
periodic boundary conditions in the space directions are imposed
on the quantum system.
Hence a quantum particle, when it is delocalized, circulates around the system.
This yields a helical flow ascending in the
imaginary-time direction of the $(d+1)$-dimensional 
imaginary-time-space. 
The pitch of the helix corresponds to the reciprocal of the
imaginary part of the energy.

It follows from the above argument that a wave function
of the periodic system acquires a complex eigenvalue as soon as it gets 
delocalized, or the corresponding flux line gets depinned and tilted.
Thus observing the energy spectrum of the non-Hermite Hamiltonian 
is a convenient way of investigating the flux-line depinning.

\makeatletter
\def\ltgt{\compoundrel<\over>}
\def\compoundrel#1\over#2{\mathpalette\compoundreL{{#1}\over{#2}}}
\def\compoundreL#1#2{\compoundREL#1#2}
\def\compoundREL#1#2\over#3{\mathrel
       {\vcenter{\hbox{$\m@th\buildrel{#1#2}\over{#1#3}$}}}}
\makeatother
Next, I explain the delocalization criterion $|\vec{g}|\ltgt\hbar\kappa$ 
in Table~\ref{tab2}.
For this purpose, I introduce the {\em imaginary} gauge 
transformation~\cite{LeDoussal}.
Suppose that we obtain an eigenfunction of the Hamiltonian for $\vec{g}=\vec{0}$
with a real energy eigenvalue:
\begin{equation}
\label{10}
{\cal H}_{0}\psi_{0}(\vec{x})=\varepsilon_{0}\psi_{0}(\vec{x}).
\end{equation}
Since the non-Hermitian field $\vec{g}$ is equivalent to an imaginary
vector potential, we may be able to gauge out the field
from the non-Hermite Hamiltonian ${\cal H}(\vec{g})$.
In other words, the eigenfunction of the Hamiltonian ${\cal H}(\vec{g})$ 
corresponding to $\psi_{0}$ may be given by
\begin{equation}
\label{20}
\psi(\vec{x})=\e^{\vec{g}\cdot\vec{x}/\hbar}\psi_{0}(\vec{x})
\end{equation}
and the eigenvalue $\varepsilon_{0}$ may remain the same.

The above transformation is valid only in a certain range of $\vec{g}$.
Assume that the wave function for $\vec{g}=\vec{0}$ is asymptotically
given by $\psi_{0}(\vec{x})\sim\e^{-\kappa|\vec{x}|}$, where $\kappa$ is
a constant.
Equation~(\ref{20}) then takes the form 
\begin{equation}
\label{30}
\psi(\vec{x})\sim\e^{-\kappa|\vec{x}|+\vec{g}\cdot\vec{x}/\hbar}.
\end{equation}
This wave function is (asymmetrically) localized for $|\vec{g}|<\hbar\kappa$.
In this region the function~(\ref{30}) satisfies periodic boundary 
conditions asymptotically in the infinite-system-size limit.
Hence the imaginary gauge transformation is valid for $|\vec{g}|<\hbar\kappa$
and Eq.~(\ref{20}) is indeed the eigenfunction of the 
non-Hermite Hamiltonian ${\cal H}(\vec{g})$ with the real eigenvalue 
$\varepsilon_{0}$.
In fact, closer inspection of the spectrum in Fig.~\ref{fig3} would reveal
that the real eigenvalues in the \lq\lq wings'' of the spectrum does not
depend on $g$.
This rigidity reflects the transverse Meissner effect of the corresponding
flux-line system.

For $|\vec{g}|>\hbar\kappa$, on the other hand, the function~(\ref{30}) 
does not satisfy periodic boundary conditions, because it blows up 
in the direction of $\vec{g}$ in this region.
Thus Eq.~(\ref{20}) is no longer an eigenfunction of the non-Hermite 
Hamiltonian ${\cal H}(\vec{g})$.
It is in this region that the eigenfunction is delocalized and acquires 
a complex eigenvalue.
(Equation~(\ref{20}) {\em is} an eigenfunction if we impose {\em open} 
boundary conditions.
All the eigenvalues remain real in this case, although the eigenfunctions
are nonetheless delocalized.
This is consistent with the fact that the helix in Fig.~\ref{fig4} never
appears if the boundaries are open in the spatial directions.)

\subsection{Mobility edge and inverse localization length}

According to the above argument, the eigenstates that belong to the 
bubble in Fig.~\ref{fig3} are delocalized, while those in the wings 
of the real eigenvalues are localized.
Hence the two vertices of the bubble, $\varepsilon_{\mathrm{c}}$ and
$-\varepsilon_{\mathrm{c}}$, are mobility edges.
The complex energy eigenvalues appear first in the middle of the energy band
because, in the one-dimensional Anderson model ($g=0$),
$\kappa$ is the smallest for the eigenstate at 
$\varepsilon=0$, as exemplified below.

Table~\ref{tab2} provides a convenient method of 
estimating the inverse localization length for $g=0$.
It is deduced from the delocalization criterion that a state at a
mobility edge for a value $g_{0}$ of the non-Hermitian field 
has the inverse localization length $\kappa=|g_{0}|/\hbar$ for $g=0$.
A numerical calculation of $\kappa$ by this method was presented
in Ref.~\cite{Hatano97} for the one-dimensional lattice model with
a box distribution.
Using the analytic result~(\ref{2006}), 
we can also calculate the inverse localization length of the 
Lloyd model~\cite{Lloyd69} (the one-dimensional
Anderson model ($g=0$) with the Lorentzian random distribution)
as a function of the energy, by solving 
\begin{equation}
\label{2007}
\varepsilon=\cosh\kappa
\sqrt{t^{2}-\gamma^{2}{\mathop{\mathrm{sech}}}^{2}\kappa},
\qquad\mbox{or}\qquad
\left(\frac{\varepsilon}{\cosh\kappa}\right)^{2}
+\left(\frac{\gamma}{\sinh\kappa}\right)^{2}=t^{2}.
\end{equation}
The solution
\begin{equation}\label{2008}
\kappa(\varepsilon)
=\cosh^{-1}\frac{\displaystyle
\sqrt{(\varepsilon+t)^{2}+\gamma^{2}}
+\sqrt{(\varepsilon-t)^{2}+\gamma^{2}}
}{2t}
\end{equation}
reproduces the exact result obtained by 
Hirota and Ishii~\cite{Hirota71,Ishii73} and Thouless~\cite{Thouless72}.

\begin{figure}
\epsfxsize=8cm
\epsfbox{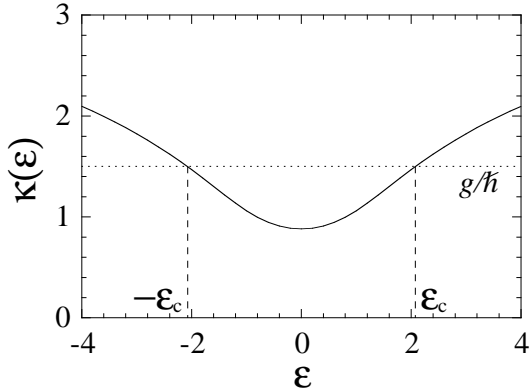}
\caption{The inverse localization length~(\protect\ref{2008}) 
of the Lloyd model (the solid line).
The parameter values used in this figure are $t=\gamma=1$.
The crossing points of the solid line and the dotted line 
(indicating a value of $g/\hbar$) yield the mobility edges
$\varepsilon_{\mathrm{c}}$ and $-\varepsilon_{\mathrm{c}}$.}
\label{fig5}
\end{figure}

Figure~\ref{fig5} shows the function $\kappa(\varepsilon)$ 
for $t=\gamma=1$.
If we apply to the system the non-Hermitian field of the value, say, 
$g/\hbar=1.5$ as indicated by a dotted line in Fig.~\ref{fig5},
the eigenstates below the dotted line 
($-\varepsilon_{\mathrm{c}}<\varepsilon<\varepsilon_{\mathrm{c}}$) 
get delocalized and form a bubble in the spectrum, 
while those above the line 
($\varepsilon<-\varepsilon_{\mathrm{c}}$ and 
$\varepsilon>\varepsilon_{\mathrm{c}}$) 
remain localized and stay in the wings of the spectrum.

\subsection{Other one-dimensional models}

Feinberg and Zee~\cite{Feinberg97c,Feinberg97d} introduced an interesting
limit of the lattice Hamiltonian~(\ref{41010}) in one dimension,
namely the one-way model:
\begin{equation}
{\cal H}\equiv 
\sum_{x}
\left(
-\left|       x+1 \right\rangle
\left\langle x \right|
+
V_{x} 
\left|       x \right\rangle
\left\langle x \right|
\right).
\end{equation}
In other words, they took an extremely non-Hermitian limit,
$t\e^{g/\hbar}\to2$ and $t\e^{-g/\hbar}\to0$.
Exact calculations of the spectral curve 
$(\mathop{\mathrm{Re}}\varepsilon,\mathop{\mathrm{Im}}\varepsilon)$
and the mobility edge $\varepsilon_{\mathrm{c}}$ become possible for various 
random distributions including the box distribution, 
the binary distribution and even diluted randomness.
See Ref.~\cite{Feinberg97d} for details.

\begin{figure}
\epsfxsize=0.495\textwidth
\begin{minipage}{0.495\textwidth}
\epsfbox{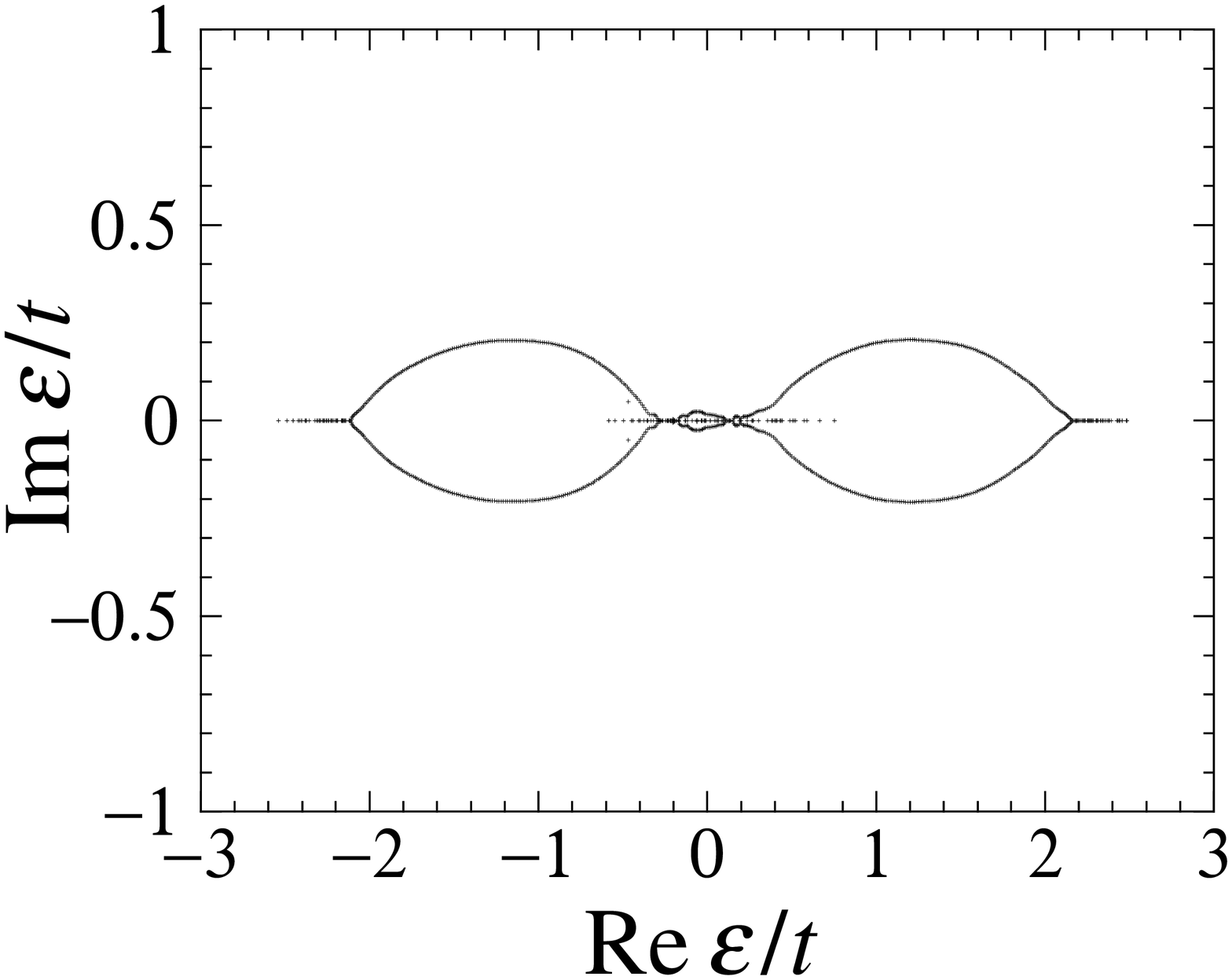}
(a)
\end{minipage}
\epsfxsize=0.495\textwidth
\begin{minipage}{0.495\textwidth}
\epsfbox{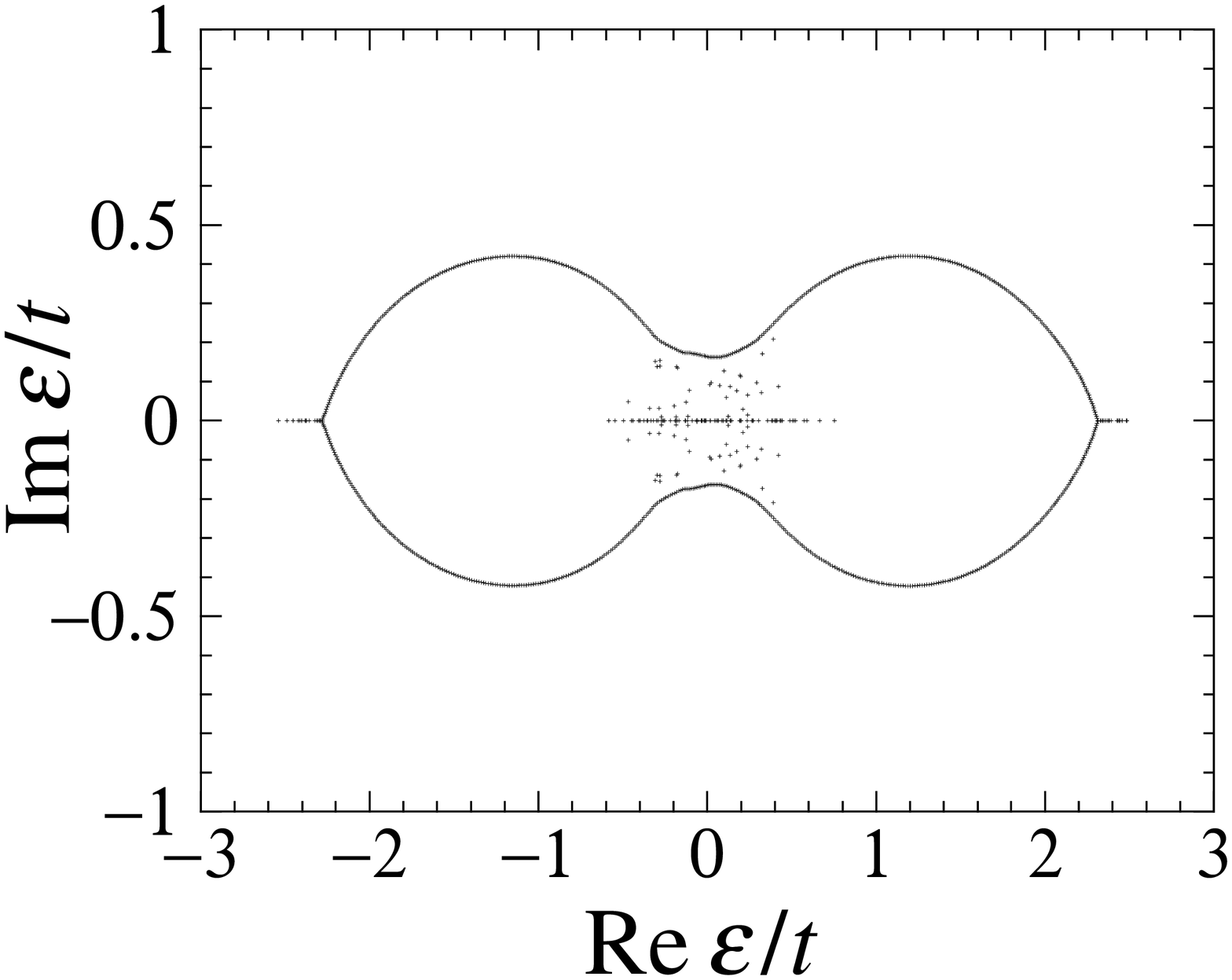}
(b)
\end{minipage}
\\
\bigskip
\epsfxsize=0.495\textwidth
\begin{minipage}{0.495\textwidth}
\epsfbox{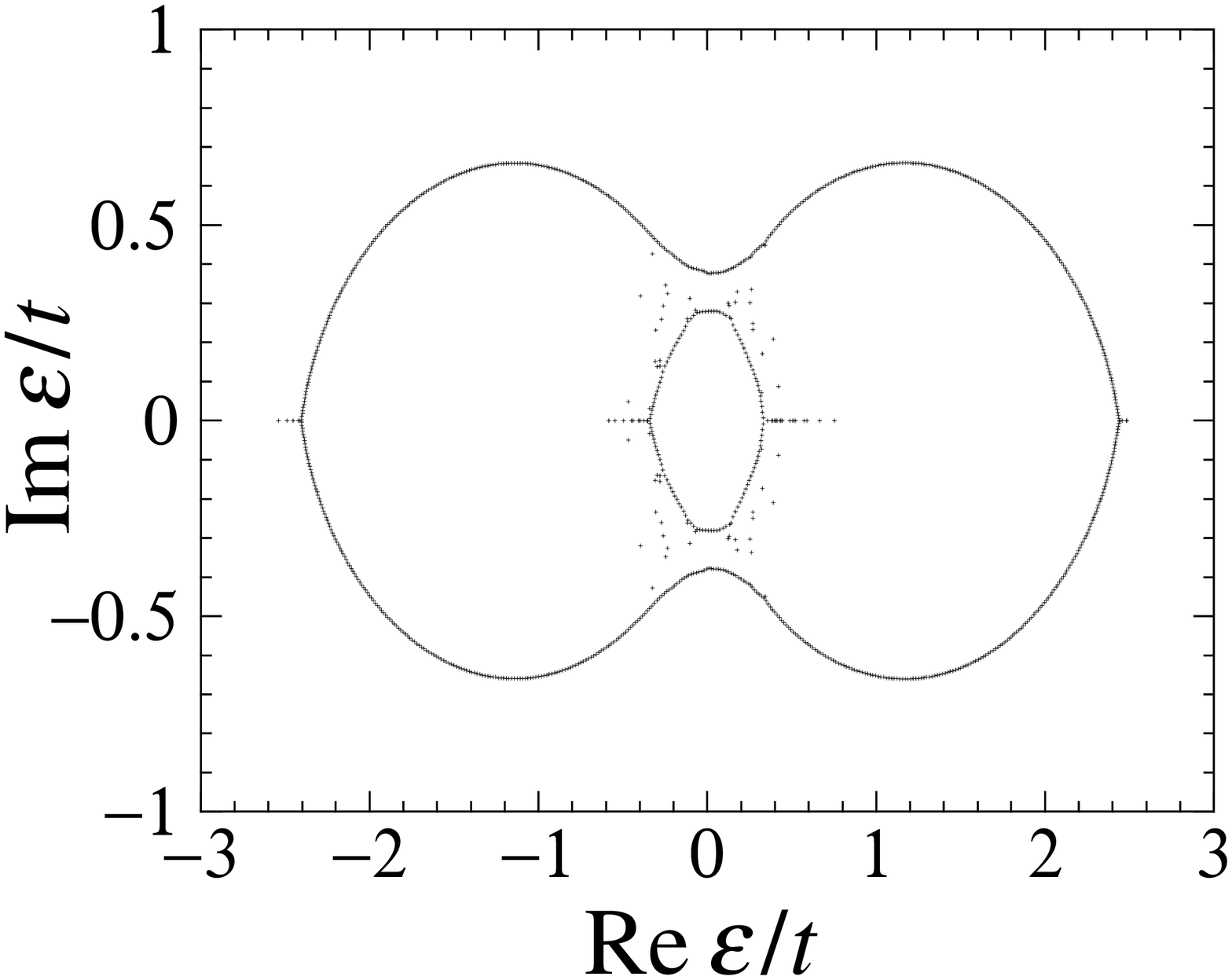}
(c)
\end{minipage}
\epsfxsize=0.495\textwidth
\begin{minipage}{0.495\textwidth}
\epsfbox{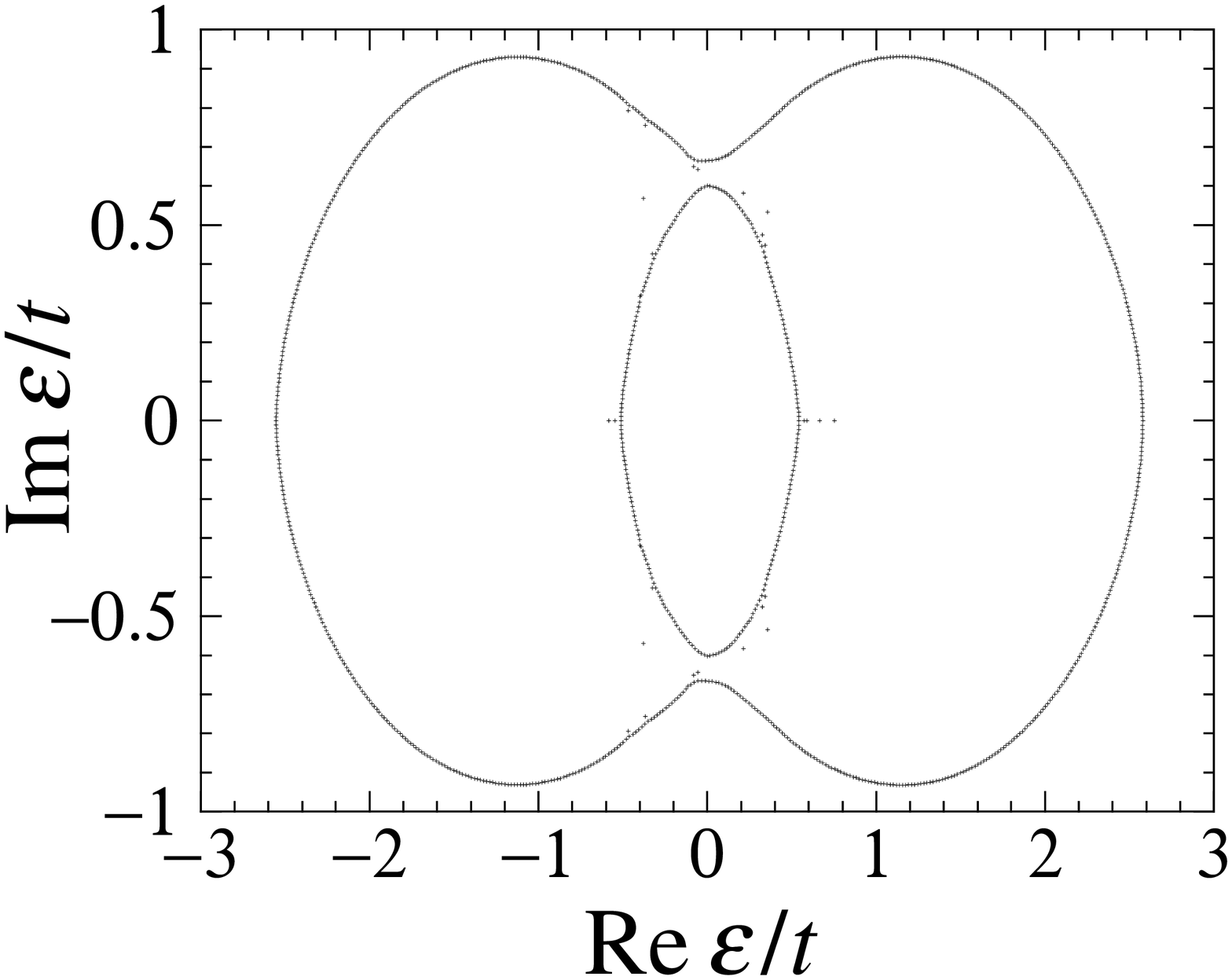}
(d)
\end{minipage}
\caption{The energy spectrum of the non-Hermitian ladder
model~(\protect\ref{ladder}) with 500$\times$2 sites.
Each eigenstate $\varepsilon$ is marked by a tiny cross 
in the complex energy plane.
The random potential at each site was chosen from a 
box distribution over the range $[-t,t]$.
The same realization of the random potential $\{V_{x}\}$ 
was used for all plots here: (a)
$g/\hbar=0.3$, (b) $g/\hbar=0.5$, (c) $g/\hbar=0.7$ and (d) $g/\hbar=0.9$.
}
\label{fig6}
\end{figure}

In Figure~\ref{fig6}, I show a numerical result of
the energy spectrum of a ladder system,
\begin{equation}\label{ladder}
{\cal H}\equiv {\cal H}_{1}+{\cal H}_{2}
-\frac{t}{2}
\sum_{x} \left(
\left|       x,2 \right\rangle
\left\langle x,1 \right|
+
\left|       x,1 \right\rangle
\left\langle x,2 \right|
\right),
\end{equation}
where ${\cal H}_{1}$ and ${\cal H}_{2}$ is the non-Hermitian Anderson 
Hamiltonian for each leg of the ladder while the last term denotes 
hopping between the legs;
\lq\lq$x,1$'' and \lq\lq$x,2$'' denote a site on the first and second leg, 
respectively.
Alternatively we may regard the labels 1 and 2 as an additional degree of 
freedom such as spin or flavor.
On the basis of the argument in the previous subsection, 
we can deduce from the result in Fig.~\ref{fig6} that there are two minima 
of $\kappa(\varepsilon)$ in the Hermitian ladder system.

The above numerical calculation was immediately followed by
Zee's analytic calculation of the spectrum for the Lorentzian
randomness~\cite{Zee97}.
Just as in the one-dimensional case~(\ref{2006}), the averaged spectrum
is squeezed towards the real axis as the randomness is increased.
Thereby we can exactly calculate $\kappa(\varepsilon)$
for the ladder Lloyd model.
The result is simply superposition of two functions of the form~(\ref{2008}).
This is consistent with the above numerical result for the box
distribution.

Note that, except in the case of the single chain, 
$\kappa(\varepsilon)$ thus calculated is an upper bound of what is referred
to as the \lq\lq inverse localization length'' in the context of the 
Anderson localization~\cite{Johnston83,Thouless83,Rodrigues86}.
The geometric average of the Green's function is taken
in defining the Anderson-localization length,
while the arithmetic average is taken in the above analytic calculation.
(The argument that yielded Table~\ref{tab2} is still valid for {\em each} 
realization of the random potential.)

\subsection{Energy spectrum in two dimension}

Finite-size calculations of the energy spectrum of the non-Hermite
Hamiltonian~(\ref{41010}) for $d=2$~\cite{Hatano96,Hatano97}
appeared to suggest the following three regions of the
non-Hermitian field $\vec{g}$. 
First, it is widely accepted for $\vec{g}=\vec{0}$ 
that all eigenstates of 
two-dimensional random systems are localized with finite localization lengths. 
Hence, by using the imaginary gauge transformation again, 
we can conclude that there is a finite region of small $\vec{g}$ where
all states remain localized. 
This is consistent with the finite-size data~\cite{Hatano96,Hatano97}.
As $\vec{g}$ is increased, delocalized states
with complex eigenvalues appear as in the one-dimensional case. 
For intermediate values of $\vec{g}$, however, 
the energy spectrum shows much more
complicated structure than in the one-dimensional case; see Ref.~\cite{Hatano97} 
for details. 
For larger $\vec{g}$, the spectrum becomes similar
to the one of the system without impurities, as was the case in $d=1$.

Nelson and Shnerb~\cite{Nelson97} suggested that the third
region of $\vec{g}$ disappears in the thermodynamic limit. 
For a very large $\vec{g}$,
or a very large transverse magnetic field $\vec{H}_{\perp}$, 
the flux line lies nearly sideways in the superconductor. 
When we project the flux line and the columnar defects onto a 
two-dimensional plane from above, the problem is approximately reduced 
to a string lying on a plane with point impurities. 
Nelson and Shnerb analyzed this reduced problem in terms of Burger's 
equation with noise~\cite{Forster77}
(or the Kardar-Parisi-Zhang equation~\cite{Kardar86})
and predicted a fractal geometry of the flux-line configuration. 
This fractality may cause the complicated energy spectrum observed 
in the second region of $\vec{g}$.
The fractal geometry does not emerge until the system size becomes 
larger than a certain crossover length, which can explain the appearance 
of the third region of $\vec{g}$
in the finite-size data~\cite{Hatano96,Hatano97}.
The above theoretical prediction is consistent with numerical results 
in Ref.~\cite{Chen96}, although the numerical estimate of the exponent 
characterizing the fractal geometry is somewhat different from the 
theoretical value predicted by Nelson and Shnerb~\cite{Nelson97}.

Note, however, that a different conclusion may be suggested by
Zee's analytic calculation of the two-dimensional spectrum
for the Lorentzian random potential~\cite{Zee97}.
In the case of the Lorentzian random distribution,
just as in one dimension, the averaged spectrum 
is squeezed towards the real axis as the randomness is increased.
In the process, the spectrum keeps the regular structure of the 
spectrum of the non-random system.

\section{Transverse Meissner effect}
\label{Meissner}

One of the interesting issues in the context of the flux-line depinning
is how the transverse Meissner effect breaks down as the 
depinning point is approached from the side of the pinning phase.
As is shown schematically in Fig.~\ref{fig2}, the pinning breaks 
first near the surface.
This is also observed in numerical results~\cite{Hatano97}.
Thus we can define a penetration depth $\tau^{\ast}$ for the 
transverse Meissner effect as the typical thickness of the sub-surface
region where the flux line is deflected from the pinning center.
(Note that $\tau^{\ast}$ is different from the penetration depth 
of the underlying superconductivity.)
The penetration depth $\tau^{\ast}$ diverges with a certain exponent,
as we increase $\vec{H}_{\perp}$, or $\vec{g}$.
This divergence leads to the breakdown of the bulk pinning.
The conclusion of Refs.~\cite{Hatano96,Hatano97} is 
\begin{equation}\label{3000}
\tau^{\ast}\sim(H_{\perp\mathrm{c}}-H_{\perp})^{-d},
\end{equation}
where $H_{\perp\mathrm{c}}$ is the depinning field.
Note that the dimensionality of the superconductor is $d+1$.

Since we regard the flux line as the world line of a quantum particle
in the framework of the non-Hermitian quantum mechanics,
the deflection of the flux line is interpreted as hopping of the 
quantum particle from a strong impurity (the localization center) 
to weaker impurities.
Near the delocalization point $g_{\mathrm{c}}$($\propto H_{\perp\mathrm{c}}$), 
Mott's argument of 
variable-range hopping~\cite{Shklovskii84} is readily applicable even to 
the non-Hermitian case.
Minimizing a hopping matrix element near the surface leads to 
Eq.~(\ref{3000}).

\section{Summary}

A class of random non-Hermitian quantum systems has been found to be 
relevant to various physical systems and to show intriguing properties.
In particular, the delocalization phenomenon of the random non-Hermitian
system is equivalent to the depinning of a flux line in type-II
superconductors.
We can investigate the delocalization just by observing the energy
spectrum of the non-Hermite Hamiltonian.
Various depinning phenomena may be understood in a similar way.
Interesting future problems include 
many-band non-Hermitian Anderson models,
detection of the fractality in the spectrum of the two-dimensional system, 
and generalization of the theory to the case of interacting systems.

\section*{Acknowledgments}
The author expresses his sincere gratitude to Prof.~D.R.~Nelson for 
collaboration and continuous encouragement.
He is also grateful to Prof.~A.~Zee for valuable and stimulating
discussions.

\appendix
\section{Path-integral mapping}
I here show derivation of Eq.~(\ref{11020-1}) from Eq.~(\ref{11020-05}).
There are various ways of formulating the path-integral mapping.
The following is based on the formulation in Ref.~\cite{Schulman81},
which employs the Trotter decomposition~\cite{Suzuki76}:
\begin{equation}
\e^{-L_{\tau}{\cal H}/\hbar}=\lim_{n\to\infty}
\left(
\e^{-{\cal V}\Delta\tau/\hbar}
\e^{-{\cal K}\Delta\tau/\hbar}
\right)^{n}.
\end{equation}
Here ${\cal K}$ and ${\cal V}$ denote the kinetic and potential terms
of the Hamiltonian~(\ref{12020}), respectively, 
and $\Delta\tau\equiv L_{\tau}/n$.

The matrix element~(\ref{11020-05}) is transformed as follows:
\begin{eqnarray}\label{pathint}
{\cal Z}&=&\lim_{n\to\infty}
\left\langle\psi^{\mathrm{f}}\Bigm|
\left(
\e^{-{\cal V}\Delta\tau/\hbar}
\e^{-{\cal K}\Delta\tau/\hbar}
\right)^{n}
\Bigm|
\psi^{\mathrm{i}}\right\rangle
\nonumber\\
&=&\lim_{n\to\infty}
\int\prod_{k=0}^{n}\d \vec{x}_{k}
\int\prod_{k=1}^{n}\d \vec{p}_{k}
\left\langle\psi^{\mathrm{f}}\Bigm|\vec{x}_{n}\right\rangle
\nonumber\\
&&\times\prod_{k=1}^{n}\left\{
\exp\left[-\frac{\Delta\tau}{\hbar}V(\vec{x}_{k})\right]
\left\langle\vec{x}_{k}\Bigm|\vec{p}_{k}\right\rangle
\exp\left[-\frac{\Delta\tau}{2m\hbar}
(\vec{p}_{k}+\mathrm{i}\vec{g})^{2}\right]
\left\langle\vec{p}_{k}\Bigm|\vec{x}_{k-1}\right\rangle
\right\}
\nonumber\\
&&\times
\left\langle\vec{x}_{0}\Bigm|\psi^{\mathrm{i}}\right\rangle.
\end{eqnarray}
The plane wave is given by
\begin{equation}
\left\langle\vec{x}_{k}\Bigm|\vec{p}_{k}\right\rangle
=\frac{1}{(2\pi\hbar)^{d/2}}
\exp\left[\frac{\mathrm{i}}{\hbar}\vec{p}_{k}\cdot\vec{x}_{k}\right]
\end{equation}
Hence the integral over each $\vec{p}_{k}$ in Eq.~(\ref{pathint})
is a Gaussian integral,
\begin{eqnarray}
\lefteqn{
\frac{1}{(2\pi\hbar)^{d}}\int\d\vec{p}_{k}
\exp\left[-\frac{\Delta\tau}{2m\hbar}(\vec{p}_{k}+\mathrm{i}\vec{g})^{2}
+\frac{\mathrm{i}}{\hbar}\vec{p}_{k}\cdot\Delta\vec{x}_{k}\right]
}\nonumber\\
&=&\left(\frac{m}{2\pi\hbar\Delta\tau}\right)^{d/2}
\exp\left[
    \frac{1}{\hbar}\vec{g}\cdot\Delta\vec{x}_{k}
    -\frac{m}{2\hbar\Delta\tau}(\Delta\vec{x}_{k})^{2}\right],
\end{eqnarray}
where $\Delta\vec{x}_{k}\equiv\vec{x}_{k}-\vec{x}_{k-1}$.
Thus Eq.~(\ref{pathint}) is reduced to
\begin{eqnarray}
{\cal Z}&=&\lim_{n\to\infty}
\left(\frac{m}{2\pi\hbar\Delta\tau}\right)^{nd/2}
\int\left(\prod_{k=0}^{n}\d \vec{x}_{k}\right)
\psi^{\mathrm{f}}(\vec{x}_{n})^{\ast}
\psi^{\mathrm{i}}(\vec{x}_{0})
\nonumber\\
&&\times
\exp\left\{
-\frac{\Delta\tau}{\hbar}\sum_{k=1}^{n}\left[
\frac{m}{2}\left(\frac{\Delta\vec{x}_{k}}{\Delta\tau}\right)^{2}
-\vec{g}\cdot\frac{\Delta\vec{x}_{k}}{\Delta\tau}
+V(\vec{x}_{k})
\right]\right\}
\nonumber\\
&=&
\int{\cal D}\vec{x}
\psi^{\mathrm{f}}(\vec{x}(L_{\tau}))^{\ast}
\psi^{\mathrm{i}}(\vec{x}(0))
\nonumber\\
&&\times
\exp\left\{
-\frac{1}{\hbar}\int_{0}^{L_{\tau}}\d\tau\left[
\frac{m}{2}\left(\frac{\d\vec{x}(\tau)}{\d\tau}\right)^{2}
-\vec{g}\cdot\frac{\d\vec{x}(\tau)}{\d\tau}
+V(\vec{x}(\tau))
\right]\right\}.
\end{eqnarray}
By assuming the free boundary conditions
$\psi^{\mathrm{i}}(\vec{x})=\psi^{\mathrm{f}}(\vec{x})=\mathrm{const.}$,
we arrive at Eq.~(\ref{11020-1}) with Eq.~(\ref{00000}).


\begin{thebibliography}{99}

\bibitem{Hatano96}
N. Hatano and D. R. Nelson,
Phys.\ Rev.\ Lett.\ 77 (1996) 570.

\bibitem{Hatano97}
N. Hatano and D. R. Nelson,
Phys.\ Rev.\ B 56 (1997) 8651.

\bibitem{Shnerb97}
N. Shnerb,
Phys.\ Rev.\ B 55 (1997) R3382.

\bibitem{Efetov97}
K.B. Efetov,
Phys.\ Rev.\ Lett.\ 79 (1997) 491.

\bibitem{Feinberg97a}
J. Feinberg and A. Zee,
Nucl.\ Phys.\ B 504 (1997) 579.

\bibitem{Janik97c}
R.A. Janik, M.A. Nowak, G. Papp and I. Zahed,
cond-mat/9705098.

\bibitem{Brouwer97}
P.W. Brouwer, P.G. Silvestov and C.W.J. Beenakker,
Phys.\ Rev.\ B 56 (1997) R4333.

\bibitem{Feinberg97c}
J. Feinberg and A. Zee,
Phys.\ Rev.\ E, to be published (cond-mat/9706218).

\bibitem{Goldsheid97}
I.Ya. Goldsheid and B.A. Khoruzhenko,
cond-mat/9707230.

\bibitem{Brezin97}
E. Brezin and A. Zee,
Nucl.\ Phys.\ B, to be published (cond-mat/9708029).

\bibitem{Nelson97}
D.R. Nelson and N. Shnerb,
cond-mat/9708071.

\bibitem{Feinberg97d}
J. Feinberg and A. Zee,
cond-mat/9710040.

\bibitem{Janik97e}
R.A. Janik, M.A. Nowak, G. Papp and I. Zahed,
hep-ph/9710103.

\bibitem{Zee97}
A. Zee,
in the same proceedings.

\bibitem{Suzuki76}
M. Suzuki,
Prog.\ Theor.\ Phys.\ 56 (1976) 1454.

\bibitem{McCoy68}
B.M. McCoy and T.T. Wu,
Il Nuovo Cimento 56B (1968) 311;
see also E.H. Lieb and F.Y. Wu,
in: Phase Transitions and Critical Phenomena Vol.\ 1,
C. Domb and M.S. Green, eds.\ (Academic Press, London, 1972) p.\ 331.

\bibitem{Miller96}
J. Miller and J. Wang,
Phys.\ Rev.\ Lett.\ 76 (1996) 1461.

\bibitem{Chalker97}
J.T. Chalker and Z.J. Wang,
Phys.\ Rev.\ Lett.\ 79 (1997) 1797.

\bibitem{Fyodorov96}
Ya.V. Fyodorov and H.-J\"{u}. Sommers,
Pis'ma Zh.\ \'{E}ksp.\ Teor.\ Fiz.\ 63 (1996) 970
[JETP Lett.\ 63 (1996) 1026].

\bibitem{Fyodorov97a}
Ya.V. Fyodorov and H.-J\"{u}. Sommers,
J.\ Math.\ Phys.\ 38 (1997) 1918.

\bibitem{Fyodorov97b}
Ya.V. Fyodorov, B.A. Khoruzhenko and H.-J\"{u}. Sommers,
Phys.\ Lett.\ A 226 (1997) 46.

\bibitem{Fyodorov97c}
Ya.V. Fyodorov, B.A. Khoruzhenko and H.-J\"{u}. Sommers,
Phys.\ Rev.\ Lett.\ 79 (1997) 557.

\bibitem{Janik97a}
R.A. Janik, M.A. Nowak, G. Papp, J. Wambach and I. Zahed,
Phys.\ Rev.\ E 55 (1997) 4100.

\bibitem{Janik97b}
R.A. Janik, M.A. Nowak, G. Papp and I. Zahed,
Nucl.\ Phys.\ B 501 (1997) 603.

\bibitem{Janik97d}
R.A. Janik, M.A. Nowak, G. Papp and I. Zahed,
hep-ph/9708418.

\bibitem{Feinberg97b}
J. Feinberg and A. Zee,
Nucl.\ Phys.\ B 501 (1997) 643.

\bibitem{Feynman65}
R.P. Feynman and A.R. Hibbs,
{\it Quantum mechanics and path integrals}
(McGraw-Hill, New York, 1965).

\bibitem{Schulman81}
L.S. Schulman,
{\it Techniques and applications of path integration}
(John Wiley \& Sons, New York, 1981) Sections~1 and~4.

\bibitem{Nelson93}
D.R. Nelson and V. Vinokur,
Phys.\ Rev.\ B 48 (1993) 13060.

\bibitem{Civale91}
L. Civale, A.D. Marwick, T.K. Worthington, M.A. Kirk, J.R. Thompson,
L. Krusin-Elbaum, Y. Sun, J.R. Clem, and F. Holtzberg,
Phys.\ Rev.\ Lett.\ 67 (1991) 648.

\bibitem{Budhani92}
R.C. Budhani, M. Suenaga, and S.H. Liou,
Phys.\ Rev.\ Lett.\ 69 (1992) 3816.

\bibitem{Chen96}
L.-W. Chen, L. Balents, M.P.A. Fisher and M.C. Marchetti,
Phys.\ Rev.\ B 54 (1996) 12798.

\bibitem{LeDoussal}
P. Le Doussal,
unpublished.
See also Sec.~IV D of Ref.~\protect\cite{Nelson93}.

\bibitem{Lloyd69}
P. Lloyd,
J.\ Phys.\ C: Solid State Phys.\ 2 (1969) 1717.

\bibitem{Hirota71}
T. Hirota and K. Ishii,
Prog.\ Theor.\ Phys.\ 45 (1971) 1713.

\bibitem{Ishii73}
K. Ishii,
Prog.\ Theor.\ Phys.\ Supple.\ 53 (1973) 77.

\bibitem{Thouless72}
D.J. Thouless,
J.\ Phys.\ C: Solid State Phys.\ 5 (1972) 77.

\bibitem{Johnston83}
R. Johnston and H. Kunz,
J.\ Phys.\ C: Solid State Phys.\ 16 (1983) 4565.

\bibitem{Thouless83}
D.J. Thouless,
J.\ Phys.\ C: Solid State Phys.\ 16 (1983) L929.

\bibitem{Rodrigues86}
D.E. Rodrigues, H.M. Pastawski adn J.F. Weisz,
Phys.\ Rev.\ B 34 (1986) 8545.

\bibitem{Forster77}
D. Forster, D.R. Nelson and M. Stephen,
Phys.\ Rev.\ A 16 (1977) 732.

\bibitem{Kardar86}
M. Kardar, G. Parisi and Y.C. Zhang,
Phys.\ Rev.\ Lett.\ 56 (1986) 889.

\bibitem{Shklovskii84}
For example, B.I. Shklovskii and A.L. Efros,
Electronic Properties of Doped Semiconductors 
(Springer-Verlag, New York, 1984).

\end{thebibliography}
\end{document}